\documentclass{elsart} 
\usepackage{graphicx}
\usepackage{amssymb}

\begin{document}
\begin{frontmatter}
\title{Calculation of the work function with a local basis set}
\author{K. Doll%\corauthref{cor}
}
\ead{k.doll@tu-bs.de} 
\address{Institut f\"ur Mathematische Physik, TU Braunschweig,
Mendelssohnstra{\ss}e 3, D-38106 Braunschweig, Germany}
%\corauth[cor]{Corresponding author.}
%
%
%
\begin{abstract}
Electronic structure codes usually allow to calculate the
work function as a part of the
theoretical description of surfaces and processes such as adsorption
thereon. This requires
a proper calculation of the electrostatic potential in all regions
of space, which is apparently straightforward to achieve with plane wave
basis sets, but more difficult with local basis sets. 
To overcome this, a relatively simple scheme is proposed to accurately
compute the work function when a local basis set is used, by having
some additional basis functions in the vacuum. Tests on various
surfaces demonstrate that a very good agreement with experimental and 
other theoretical data can be achieved.

\end{abstract}
\end{frontmatter}
%
% main text
%
\section{Introduction}

The work function is one of the most important quantities in the
characterization of surfaces and processes on surfaces such as adsorption
or chemical reactions. It is defined as the energy required to move an electron
from deep within the bulk to a point far away from the surface. 
Experimentally,
'far away' means a distance large compared to atomic dimensions, but
small compared to the dimensions of the corresponding face of the
crystal; the size of the sample is finite. In electronic
structure codes, surfaces can be modeled as slabs having an infinite extension
in the surface plane, and a finite thickness orthogonal to it.
The work function is then obtained as the difference
of the energy of an electron at infinity, minus the Fermi energy:
$E(\infty)-E_F$. 

The calculation of the work function is nowadays routinely done in density
functional calculations, and the agreement between theoretical and
experimental data is usually very good, see e.g. 
\cite{SkriverPRBp7157}. To compute $E(\infty)-E_F$, knowledge
of the electrostatic potential, which is determined by the charge density,
and of the Fermi energy is necessary. If the
exact functional was known, and if numerical noise was neglected, then
the charge density, the position of the Fermi energy and thus the 
work function would be obtained exactly
\cite{Sham1966,Janak,PerdewNATO}.

Most codes use plane waves as basis functions, where
the description of the electrostatic potential 
in all parts of space, i.e. in the region of the surface and in
the vacuum region, seems to be without any difficulty. 
Local basis sets such as Gaussian type orbitals are an alternative
to plane waves. They perform well for many properties such as energetics
or structural optimization. However, the calculation of the
work function appears to be more intricate,
and difficulties such as a 
basis set dependence of the results had been observed 
\cite{Boettgeretal1995,nickel,anna2006}.

To overcome this problem and to obtain accurate values for the work function, 
a simple scheme is proposed. It consists of
having additional basis functions in the vacuum region above the surface,
for a better description of the electrostatic potential and thus the 
work function. The method and computational details are explained in section 
\ref{methodsection}.
In section \ref{cusection},
tests on the low index Cu surfaces are performed. 
To demonstrate the 
reliability of this scheme,
results for various clean surfaces are presented in
section  \ref{cleansurfacessection}, and for the adsorbate systems
Cl/Cu(111) and K/Ag(111) in section \ref{adsorbates}.

\par
\section{Method}
\label{methodsection}

The calculations were performed with the code {\sc CRYSTAL2003} 
\cite{Manual03}. This code uses Gaussian type orbitals, which can 
be centered at the position of the atoms. 
In addition, the code has the option
to use basis functions without atoms (usually referred to as
dummy or ghost atoms)
which is the standard procedure when
the basis set superposition error is evaluated by the counterpoise method.
In the present context, 
this option will be used for a better description of the
electrostatic potential in the vacuum region. Such an approach had
already been used earlier, e.g. for Pt(110) and Pd(110) \cite{Feibelman1995}.
A different way to tackle the problem of computing
the work function might
be to try basis sets with very diffuse exponents, and to explore
whether convergence of the work function can be achieved. However, this
can not be done with the CRYSTAL code because linear dependence sets
in when very diffuse exponents (i.e. with exponents less than $\sim$ 0.1)
are used, and the calculations become numerically unstable.
The idea is thus to calibrate a scheme relying on ghost atoms, and
to perform extensive tests on various surfaces
and adsorbate systems. 
 
The basis sets employed in the present work
are either pure all electron basis sets or use 
in addition a pseudopotential.
Using a pseudopotential is not mandatory, and calculations
on heavy atoms with all electron basis sets are feasible. 
The pseudopotential helps to reduce the computational effort, and, maybe even 
more important, offers the possibility to include scalar-relativistic
effects. The Gaussian basis sets are:
a $[3s2p]$ basis set or a $[4s3p]$ basis set for Li \cite{lithium}, 
$[6s5p2d]$ basis sets for 
Cu\cite{clcu} and Ni\cite{nickel}, a $[4s3p2d]$ basis set together with
a 19-valence electron pseudopotential for Ag \cite{silver},
a $[4s4p2d]$ basis set for Pt together with a 18-valence electron
pseudopotential \cite{pt}, 
and  $[5s4p1d]$ basis sets for Cl\cite{clcu},  
and  K \cite{KCu111}. 
$\vec k$-point nets of the size 16 $\times$ 16 were used, and the
smearing temperature was in the range between 0.001 $E_h$ and a maximum
value of 0.01 $E_h$, as described earlier \cite{lithium,clcu,silver,pt,kag111}.
In the case of nickel, spin-polarized calculations were performed,
and the smearing temperature must be chosen low (0.001 $E_h$) because
a too high temperature would artificially reduce the magnetic moment.

Most of the calculations were done at the level of the
local density approximation (LDA), with the Perdew-Zunger potential
\cite{PZ}. In some cases, where computationally
expensive optimizations had been performed earlier \cite{clcu,kag111}, 
the same
functionals were used here again:
the gradient corrected functional of Perdew and Wang (PWGGA) or
Perdew, Burke and Ernzerhof (PBE). 

\begin{figure}[ht]
\begin{center}
\includegraphics[width=3cm]{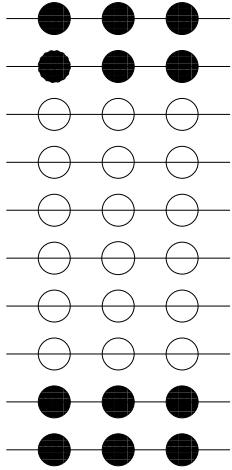}
\caption{\label{anordnungsfigur} Slab model used: there are some (here: 6)
layers made of atoms with basis set (white circles), and a few
layers where only basis functions are placed, without nuclear charge
(referred to as ghost atoms, or ghost layers;
here: 2 on each side, dark circles).}
\end{center}
\end{figure}

The surfaces are modeled with slabs of a finite thickness (typically
6 layers for the clean surfaces and 3-5 adsorbate layers in the case of
adsorbate systems), as displayed in Fig. \ref{anordnungsfigur}. 
The slabs are not periodically repeated in the third
dimension. The slab model should be thick
enough so that the Fermi energy is not modified by surface states.
The zero of the electrostatic potential $\Phi$ is defined by the
CRYSTAL code in such a way that $\Phi(\infty)=-\Phi(-\infty)$, and the Fermi
energy is then determined by the number of electrons. In the case
of symmetrical arrangements of the slabs (e.g. clean, unrelaxed surfaces),
$\Phi(\infty)=0$ holds, and the work function is $-E_F$.

\section{The low index Cu surfaces}
\label{cusection}

The Cu(111) surface is chosen as a first system for the method suggested. 
Various tests are performed: first, relaxation is not allowed and
the distance is determined by the Cu lattice constant, as computed at the
LDA level (3.53 \AA \cite{clcu,cocu}). Three
basis sets for the ghost layers are compared: one basis
set which consists of only the outermost diffuse $sp$ shell of the
original Cu basis set (exponent
0.15), a second basis set consisting of this $sp$ shell and additionally
the outermost diffuse $d$ function (exponent 0.392), and the full
basis set, i.e. the same basis set is used for the ghost atoms and for
the Cu atoms which are not ghosts. 

Various numbers of ghost layers are tested. 
The ghost atoms are placed as if the surface was
continued in the subsequent ghost layers, i.e. like in the 
fcc lattice, symmetrically on both sides of the slab. 
This arrangement is shown in figure \ref{anordnungsfigur}.

\begin{table}[t]
\caption{\label{Cu111unrelaxed}The work function of Cu(111), in $E_h$
(1 $E_h$=27.2114 eV),
with
various layers of ghost atoms, and three basis sets for the ghost atoms, 
at the level
of the LDA. The number of ghost layers corresponds to the sum of the
layers on both sides of the slab.}
\begin{tabular}{cccc}\hline
number of & Ghost basis set:  & Ghost basis set:  
& Ghost basis set: \\
ghost layers & $sp$ 0.15 & $sp$ 0.15, $d$ 0.392 & full basis set\\ 
& work function & work function & work function\\ \hline
0 & 0.142 & 0.142 & 0.142\\ 
2 & 0.189 & 0.190 & 0.190\\
4 & 0.190 & 0.190 & 0.191\\
6 & 0.190 & 0.190 & 0.191\\
\hline
\end{tabular}
\end{table}

The data are displayed in table \ref{Cu111unrelaxed}. We note that
already two ghost layers, i.e. one on each side, result in a value
of the work function which is practically converged, as the
value hardly changes when more ghost layers are used. Another
important finding is that the outermost diffuse exponents are
sufficient for a good description. This is expected, as these
diffuse exponents are the ones which have the most impact to describe
the electrostatic potential in the vacuum region, and diffuse exponents are 
required to describe a delocalized charge distribution. \\

\begin{figure}[ht]
\begin{center}
\includegraphics[width=14cm]{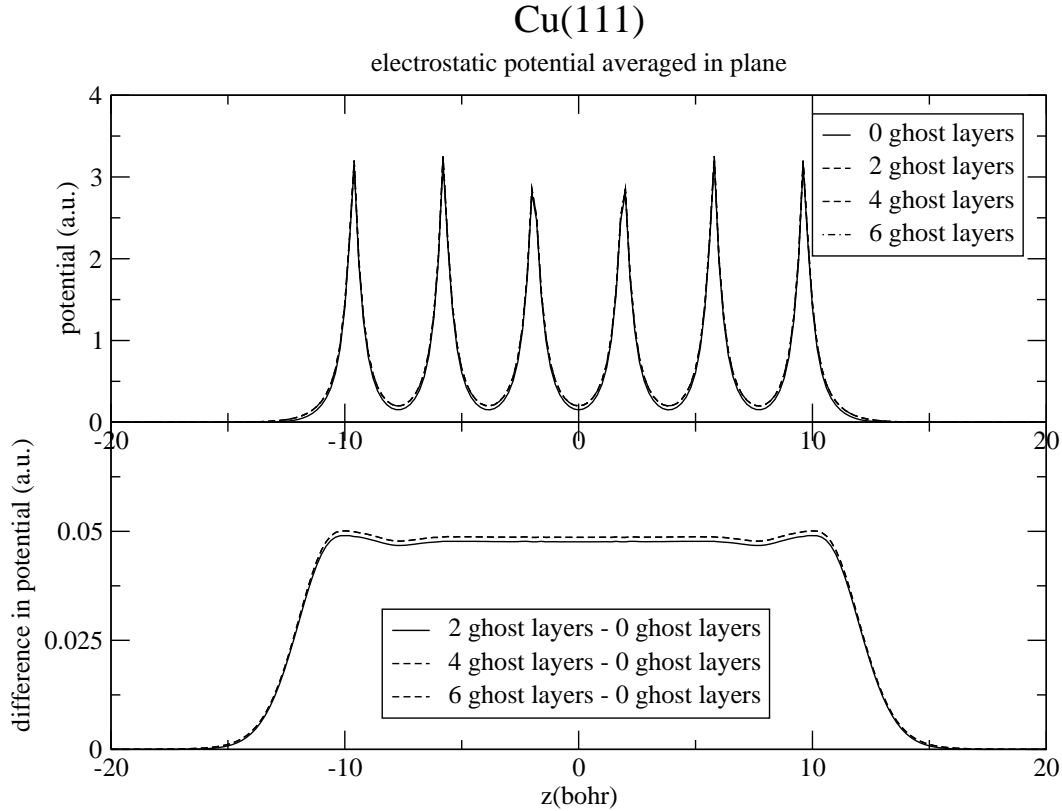}\vspace{1cm}
\caption{\label{cu111potential} 
The electrostatic potential of the Cu(111)
surface, averaged in the plane parallel to the surface, with and without
additional layers with dummy atoms (top panel). The difference of
the potential with additional layers with ghost atoms and the
potential without ghost atoms (bottom panel).}
\end{center}
\end{figure}

The electrostatic potential for the Cu(111) surface
is visualized in figure \ref{cu111potential}, with various numbers
of ghost layers (note that, to obtain the electrostatic energy of an electron,
the potential has to be multiplied by the electronic charge, -$|e|$). As
the total potential (top panel) looks very similar for all possible
arrangements, it is better to consider the difference of the electrostatic
potential with and without ghost layers (bottom panel).
This difference of the potentials with 2,4 or 6 ghost
layers and without ghost layers demonstrates that convergence with
respect to the number of ghost layers is very fast, and the change
in the potential is about 0.05 a.u. in the middle of the slab
which corresponds to the change in the work function when having
ghost layers, as displayed in table \ref{Cu111unrelaxed}.

In a next step, relaxation of the top Cu layer on each side of the slab
was allowed. One $sp$ basis function was used on the
ghost atoms, with 6 layers made of ghost atoms
(3 on each side of the Cu slab). The positions of the ghost atoms were
held fixed. 
The data in table \ref{Cu100110111} shows that in all cases,
the work function hardly changed.
The relaxation changes slightly when ghost layers are used. The ghosts
thus have a slight impact on the geometry, as they can also be interpreted
as an improvement of the basis set of the outermost atom.

\begin{table}[ht]
\caption{\label{Cu100110111}The work function of the Cu(100),
Cu(110) and Cu(111) surface, in $E_h$ (1 $E_h$=27.2114 eV),
with and without
layers of ghost atoms, at the level of the LDA. The top
Cu layer on each side is first not allowed to relax (first two
rows), and later allowed to relax (third and fourth row).}
\begin{tabular}{ccccccc}\hline
& \multicolumn{2}{c}{Cu(100)} &\multicolumn{2}{c}{Cu(110)} &
 \multicolumn{2}{c}{Cu(111)} \\
number of &  relaxation & work &  relaxation & work &  relaxation & work \\
ghost layers  & (\AA) & function & (\AA) & function & (\AA) & function \\
\hline
0 & - & 0.129 & - & 0.121 & - & 0.142 \\
6 & - & 0.180 & - & 0.173 & - & 0.190 \\ \\
0 & -0.06 & 0.129 & -0.11 & 0.123 & -0.028 & 0.142 \\
6 & -0.04 & 0.179 & -0.08 & 0.173 & -0.013 & 0.190 \\ \\
Experiment\cite{Michaelson} & & 0.169 & 
       & 0.165  & 
       & 0.183  \\ \\
Theory\cite{SkriverPRBp7157} & & 0.193  &
         & 0.165 &
         & 0.195  \\ \hline
\end{tabular}
\end{table}

\section{Further metal surfaces}
\label{cleansurfacessection}

In the following, other examples of metal surfaces are considered, at the
level of the LDA.
These are Li (with a computed equilibrium lattice constant of 3.369 \AA),
Ni (3.43 \AA), Ag (3.98 \AA), Pt (3.94 \AA). The surfaces were not relaxed. 
As basis functions for the ghosts, the outermost
diffuse basis functions of the respective basis sets were used.
The data in table
\ref{metalsgeneral} 
demonstrates that a fast convergence with respect to the number of 
ghost layers is achieved and usually
2 layers of ghost atoms (1 on each side of the slab) are already 
sufficient. The data is in good agreement with the
experimental data, as far as a comparison is possible.  
The order of magnitude of the 
deviations from other theoretical data is reasonable: for example,
in \cite{Baudetal}, by simply changing the LDA exchange
correlation potential, variations in the work function between
5.96 eV=0.219 $E_h$ 
(Vosko, Wilk and Nusair potential) and 6.57 eV=0.241 $E_h$ 
(von Barth and Hedin potential) were observed for the Pt(111) surface.
For Li, two different basis sets were used (whereas the Li ghost basis set
was fixed, $sp$=0.10). As expected, 
the larger $[4s3p]$ basis set performs better in
the case of having no or two ghost layers, but from 4 ghost layers onwards, the
$[3s2p]$ and the $[4s3p]$ basis sets give virtually identical work functions. 
Larger basis sets are practically impossible: 
for Li(110), already the $[4s3p]$ basis set leads to numerical instability.

\begin{table}[ht]
\caption{\label{metalsgeneral}The work function, in $E_h$
(1 $E_h$=27.2114 eV), of several metals, computed 
at the level of the LDA. Six substrate layers and from 0 to 6 
layers (i.e. 0 to 3 on each side) with ghost atoms were used. As indicated
in the table, two different basis sets were used for Li.}
\begin{tabular}{cccccccc}\hline
surface & & \multicolumn{6}{c}{work function} \\
        & & \multicolumn{4}{c}{number of ghost layers:} & experiment & theory\\
        & & 0 & 2 & 4 & 6 & \\ \hline
Li(100) $[3s2p]$ & & 0.083 & 0.114 & 0.119 & 0.120 & (0.107)$^a$  
\cite{Michaelson} & 0.111 
\cite{KokkoPRB1995}\\
Li(100) $[4s3p]$ & & 0.107 & 0.116 & 0.119 & 0.120 \\
Li(110) $[3s2p]$ 
& & 0.094 & 0.129 & 0.130 & 0.130 & & 0.126 \cite{KokkoPRB1995} \\
Li(111) $[3s2p]$
& & 0.073 & 0.098 & 0.111 & 0.115 & & 0.115 \cite{KokkoPRB1995}\\
Li(111) $[4s3p]$ & & 0.101 & 0.107 & 0.113 & 0.115 &  \\
Ni(100) & & 0.162 & 0.194 & 0.195 & 0.195 & 0.192 \cite{Michaelson}
& 0.183 \cite{Mittendorfer}\\
Ni(110) & & 0.146 & 0.173 & 0.180 & 0.180 & 0.185 \cite{Michaelson}
& 0.169 \cite{Mittendorfer}\\
Ni(111) & & 0.173 & 0.202 & 0.202 & 0.202 & 0.197 \cite{Michaelson}
& 0.188 \cite{Mittendorfer}\\
Ag(100) & & 0.147 & 0.177 & 0.177 & 0.177 & 0.171 \cite{Michaelson}
& 0.184
\cite{SkriverPRBp7157}\\
Ag(110) & & 0.140 & 0.167 & 0.172 & 0.172 & 0.166 \cite{Michaelson}
& 0.162
\cite{SkriverPRBp7157}\\
Ag(111) & & 0.153 & 0.183 & 0.183 & 0.182 & 0.174 \cite{Michaelson}
& 0.184 
\cite{SkriverPRBp7157}\\
Pt(100) & & 0.216 & 0.223 & 0.223 & 0.223 & (0.214)$^b$ \cite{Salmeron} 
& 0.240 \cite{Baudetal}\\
Pt(110) & & 0.203 & 0.209 & 0.211 & 0.211 & (0.215)$^b$ \cite{CRC} 
& 0.227 \cite{Baudetal}\\
Pt(111) & & 0.219 & 0.226 & 0.225 & 0.225 & 0.223 \cite{Salmeron}
& 0.240 \cite{Baudetal}\\
\hline
\end{tabular}

$^a${The experimental 
value for Li is in brackets as it refers to polycrystalline data.}
$^b${The experimental values for Pt(100) and Pt(110)
are in brackets as these surfaces reconstruct.}
\end{table}

\section{Adsorbate systems}
\label{adsorbates}

An interesting and important application of work functions are
adsorbate systems. Therefore, the scheme is now
tested with Cl/Cu(111) and K/Ag(111). Both systems had been studied
previously with a local basis set\cite{clcu,kag111}, 
and now the work functions are evaluated.
With the adsorbate adsorbed on one side of the slab (on the side pointing
to +$\infty$), $E(\infty )-E_F$
gives the work function of the adsorbate covered side.
$E(-\infty )-E_F$ corresponds to the work function of the
clean surface and is virtually identical to the values from tables
\ref{Cu100110111} and \ref{metalsgeneral}, when computed for Cl/Cu(111) or
K/Ag(111).

First, the adsorbate system Cl/Cu(111), at the coverage of one third,
in a $(\protect\sqrt{3} \times \protect\sqrt{3})$R30$^\circ$ pattern,
is considered. Chlorine is adsorbed on the fcc site
\cite{Crapper1986}, with a minority occupation of
the hcp site\cite{Kadodwala}. This was confirmed by total energy
calculations\cite{clcu}, and the geometry was in excellent agreement
with the experimental geometry. With this geometry, two ghost layers
(one on each side of the slab) are now added to compute the work function,
see table \ref{ClCu111table}. First, the work function of the clean Cu(111)
surface is determined as 0.175 $E_h$ (1 $E_h$=27.2114 eV), at the PWGGA level.
When Cl is adsorbed it is found to increase by 0.023 $E_h$ at the fcc site, 
which is reasonable, compared with computed data\cite{Migani2005}
(0.01 $E_h$ increase, for a coverage of 0.125) and experimental 
data\cite{GoddardLambert},
where an increase of the work function up to a saturation of 0.04 $E_h$
was observed (however, the coverage was not specified). The
increase of the work function is larger for the bridge and largest for
the top site, which is consistent with the increasing Mulliken
charge on chlorine and the increasing interlayer distance between
the Cl layer and the top Cu(111) layer. It is also interesting
that the chlorine Mulliken charge is $\sim$ -0.2 and thus far away from
that of a fully negatively charged chlorine ion.

In earlier calculations on nickel surfaces\cite{nickel}, without using
ghost atoms, it had already been shown
that varying the outermost diffuse exponent changed the work function
strongly, 
but the other properties such as the geometry, relative energies
of the adsorption sites or Mulliken populations were only weakly affected.
As the relative energies of the various sites are very important,
the relative energies with and without ghosts were computed, as
a further test for the system Cl/Cu(111).
These data are included in table \ref{ClCu111table}.
Essentially, the energy splitting between the various sites is the same
with or without ghost atoms, as the largest difference 
is in the range of the numerical noise (0.0155 $E_h$ without
ghosts versus 0.0147 $E_h$ with ghosts, i.e. 0.0008 $E_h$,
in the case of the top site).

\begin{table}[t]
\caption{\label{ClCu111table}
The work function of the adsorbate system 
Cu(111)$(\protect\sqrt{3} \times \protect\sqrt{3})$R30$^\circ$-Cl, at the PWGGA level; the 
interlayer distance Cl-Cu(111); the Mulliken charge of the chlorine atom;
and the binding energy per Cl atom, relative to the fcc site, with and
without ghosts.}
\begin{tabular}{cccccc}\hline
 Cl site & work function & distance Cl-Cu(111) & Cl charge & 
\multicolumn{2}{c}{binding energy, relative to fcc site} \\
 & & & & with ghosts & without ghosts \\
& ($E_h$) & (\AA) & ($|e|$) & $(E_h)$ & $(E_h)$ \\
\hline
 fcc
& 0.198 & 1.89 & -0.19 & 0 & 0 \\
 hcp
& 0.200 & 1.90 & -0.20 & 0.0001 & 0.0002 \\
 bridge
& 0.206 & 1.94 & -0.23 & 0.0024 & 0.0025 \\
 top
& 0.238 & 2.17 & -0.31 & 0.0147 & 0.0155 \\ \hline
\end{tabular} 
\end{table}

In addition, the work function of the system K/Ag(111) is computed (table 
\ref{KAg111}, at the PBE level, for the clean surface and at
the coverages of one fourth and one third of a monolayer.
With four ghost layers (two on each side
of the slab), the computed work function is in reasonable agreement with
the experimental value for the clean surface and with the data
measured at various coverages\cite{Blass}; 
the initial decrease of the work function
when potassium is adsorbed is found in theory and experiment, and similarly,
the slight increase with larger coverage and depolarization is observed.

\begin{table}[ht]
\caption{\label{KAg111}The work function and its change upon
adsorption for the adsorbate system K/Ag(111), at the PBE level.}
\begin{tabular}{ccccccc}\hline
surface & K site & \multicolumn{2}{c}{work function}  \\
 & &  this work & exp. \\
Ag(111) & & 0.164 & 0.17 \\
Ag(111)$(2 \times 2)$-K & fcc \cite{Leathermanetal} & 0.068 &  $\sim$ 0.05 \cite{Blass}
 \\
Ag(111)$(\protect\sqrt{3} \times \protect\sqrt{3})$R30$^\circ$-K 
& hcp \cite{Leathermanetal} & 0.084 &  $\sim$ 0.06 \cite{Blass}  \\ \hline
\end{tabular}
\end{table}

Finally, it should be emphasized that the approach using ghost atoms 
as displayed in Fig. \ref{anordnungsfigur}
will need further refinements in cases such as for example
low coverages or reconstructions,
where it may be necessary to have additional ghost atoms,
for instance in surface
regions not covered by adsorbates, or at the place of
missing rows.

\section{Conclusion}
A scheme to accurately compute
work functions with a local basis set was suggested and tested. 
Placing at least one layer with ghost atoms in the vacuum region on each
side of the slab makes it feasible 
to obtain reasonable values for the work function.
It is sufficient to use the outermost diffuse basis functions for
the ghost atoms.
Structural and energetical properties change only weakly when ghost
atoms are added. Computed work functions for simple metals and
for adsorbate systems are in good agreement with data obtained
with codes employing plane waves and with experimental data.

\clearpage

\bibliographystyle{elsart-num}

\end{document}